\documentclass[useAMS,usenatbib,twocolumn,preprintnumbers,nofootinbib]{revtex4}
\def \be {\begin{equation}} 
\def \ee {\end{equation}} 
\def \bea {\begin{eqnarray}} 
\def \eea {\end{eqnarray}} 

\usepackage{graphicx}
\usepackage{dcolumn}
\usepackage{bm}
\usepackage{epsfig} 
\usepackage{amsfonts}
\usepackage{amsmath}
\usepackage{amssymb}
\usepackage[usenames]{color}
\usepackage[colorlinks=true]{hyperref}

\hypersetup{colorlinks,citecolor=blue,linkcolor=blue,urlcolor=blue}
\hypersetup{final=true}

\begin{document}

\title{How does an incomplete sky coverage affect the Hubble Constant variance?}

\author{Carlos A. P. Bengaly$^{1}$\email{carlosap87@gmail.com}, Uendert Andrade$^{2}$\email{uendertandrade@on.br}, Jailson S. Alcaniz$^{2,3}$\email{alcaniz@on.br}}
\affiliation{$^1$Department of Physics \& Astronomy, University of the Western Cape, Cape Town 7535, South Africa}
\affiliation{$^2$Observat\'orio Nacional, 20921-400, Rio de Janeiro - RJ, Brasil}
\affiliation{$^3$Departamento de F\'{\i}sica, Universidade Federal do Rio Grande do Norte, 59072-970, Natal, RN, Brasil}

\date{\today}

\begin{abstract}
We address the $\simeq 4.4\sigma$ tension between local and the CMB measurements of the Hubble Constant using simulated Type Ia Supernova (SN) data-sets. We probe its directional dependence by means of a hemispherical comparison through the entire celestial sphere as an estimator of the $H_0$ cosmic variance. We perform Monte Carlo simulations assuming isotropic and non-uniform distributions of data points, the latter coinciding with the real data. This allows us to incorporate observational features, such as the sample incompleteness, in our estimation. We obtain that this tension can be alleviated to $3.4\sigma$ for isotropic realizations, and $2.7\sigma$ for non-uniform ones. We also find that the $H_0$ variance is largely reduced if the data-sets are augmented to 4 and 10 times the current size. Future surveys will be able to tell whether the Hubble Constant tension happens due to unaccounted cosmic variance, or whether it is an actual indication of physics beyond the standard cosmological model. 
\end{abstract}

\maketitle

\section{Introduction}\label{sec:intro}

Precise measurements of the Hubble Constant ($H_0 = 100h_0$ $\rm{km \, s^{-1} \, Mpc^{-1}}$) are of great interest not only to improve our  understanding of the current cosmic evolution but also to provide new insights into some fundamental questions of the physics of the early Universe (see e.g.,~\cite{suyu12} for a broad discussion). Currently, there is a $\sim 4.4\sigma$ tension between measurements of $h_0$ performed in the local universe using type Ia supernovae (SN) observations calibrated with Cepheid distances to SN host galaxies,
$H_0 = 74.03 \pm 1.42 \, \rm{m \, s^{-1} \, Mpc^{-1}}$ (hereafter R19~\citep{riess19}; see also~\cite{riess16, riess18a}), and the estimates of $H_0$ obtained from the Cosmic Microwave Background (CMB) data in the context of the standard $\Lambda$CDM model, $H_0 = 67.36 \pm 0.54 \, \rm{km \, s^{-1} \, Mpc^{-1}}$ (hereafter P18~\citep{planck18}; see also~\citep{planck16a}). Many previous works attempted to address this problem using different approaches, e.g., in light of the cosmic variance due to nearby inhomogeneities~\citep{marra13, wojtak14, bendayan14, odderskov14, bengaly16, odderskov16, odderskov17, chiang17, romano18, macpherson18, kenworthy19, lukovic19, boehringer19}, re-calibrating the distance ladder~\citep{efstathiou14, cardona17, feeney18, follin18, wu18, shanks18, riess18b, taubenberger19, camarena19, lombriser19}, or looking at extensions to the standard model of Cosmology~\citep{bernal16, benetti17, divalentino18, benetti18, bolejko18, mortsell18, camarena18, graef19, poulin19, vattis19, agrawal19, adhikari19, carneiro19, vagnozzi19}. Meta-studies of the $H_0$ measurements and tension also confirmed a $H_0$ tension above $3\sigma$~\cite{crossland19}. New Hubble Constant measurements from strongly lensed quasars time delays~\cite{wong19}, and using the tip of the red giant branch~\cite{freedman19, yuan19}, could not solve this issue~\cite{verde19}. 

Our goal in this paper is to quantify how the incompleteness of current SN data can affect the $H_0$ variance, i.e., a shot noise variance estimator. Rather than reassessing these measurements, we focus on quantifying how does the tension between the $H_0$ measurements change due to it. To do this, we map the anisotropy of the Hubble Constant across the sky using low-$z$ SN ($0.023 < z  < 0.150$) from simulations based on the latest compilation available, namely, the Pantheon data~\citep{scolnic18}. We produce sets of Monte Carlo (MC) realizations assuming different sky coverage configurations which coincide with isotropic and the real data (non-uniform) sky distributions. This allows to assess how the SN sampling due to the uneven redshift distribution, {in addition to} their distance uncertainties and incomplete celestial distribution of data points, impacts the $H_0$ measurements. We also discuss how much does the $H_0$ cosmic variance decrease when larger, more homogeneous SN data sets are considered. If the cosmic variance becomes much lower than the current values, we can rule it out as a possible explanation for this result, thereby strengthening the hypothesis of physics beyond the concordance model.

\section{Data Analysis}\label{sec:DA} 

\begin{figure*}
\includegraphics[width=0.32\textwidth, height=0.215\textheight]{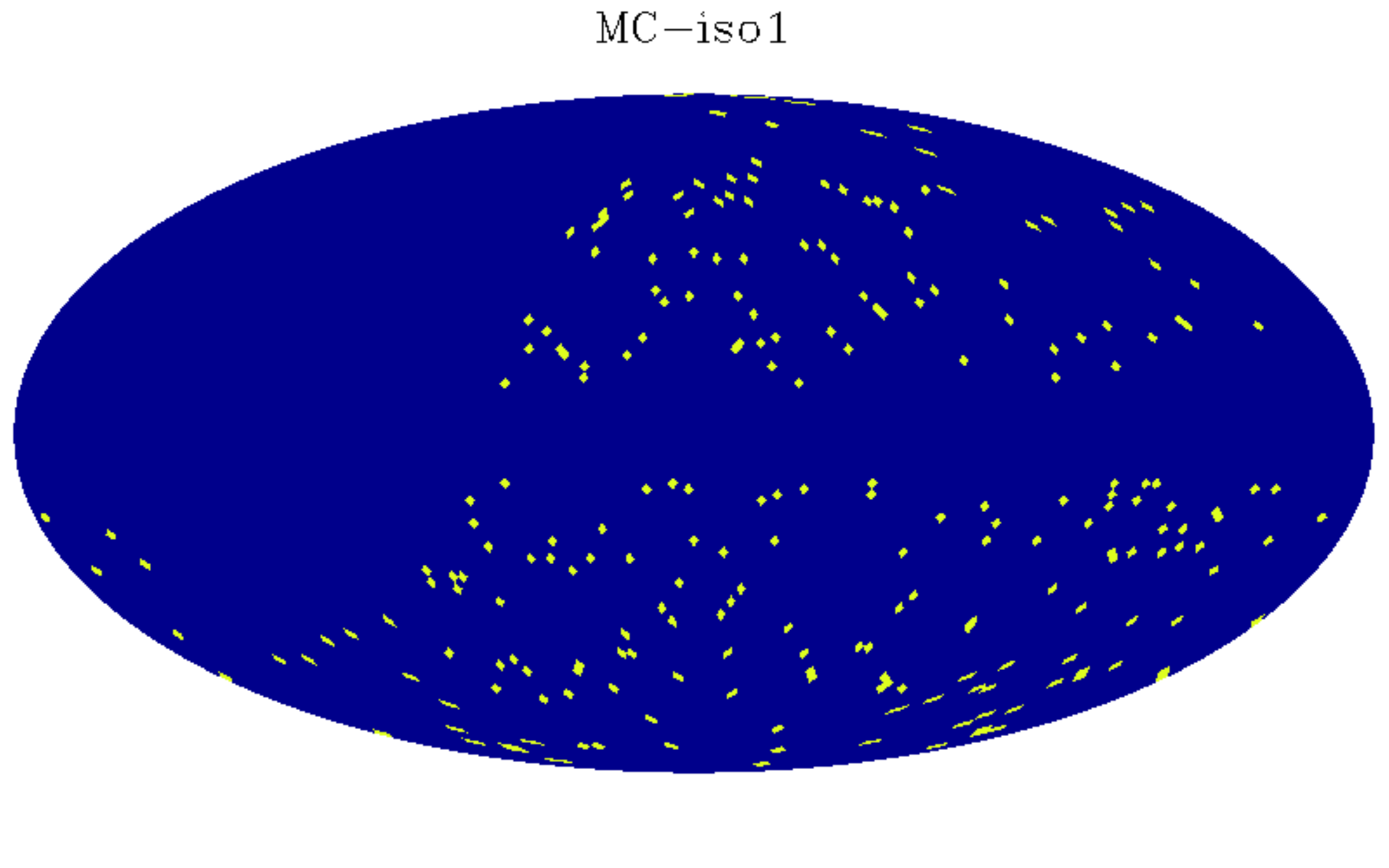}
\includegraphics[width=0.32\textwidth, height=0.215\textheight]{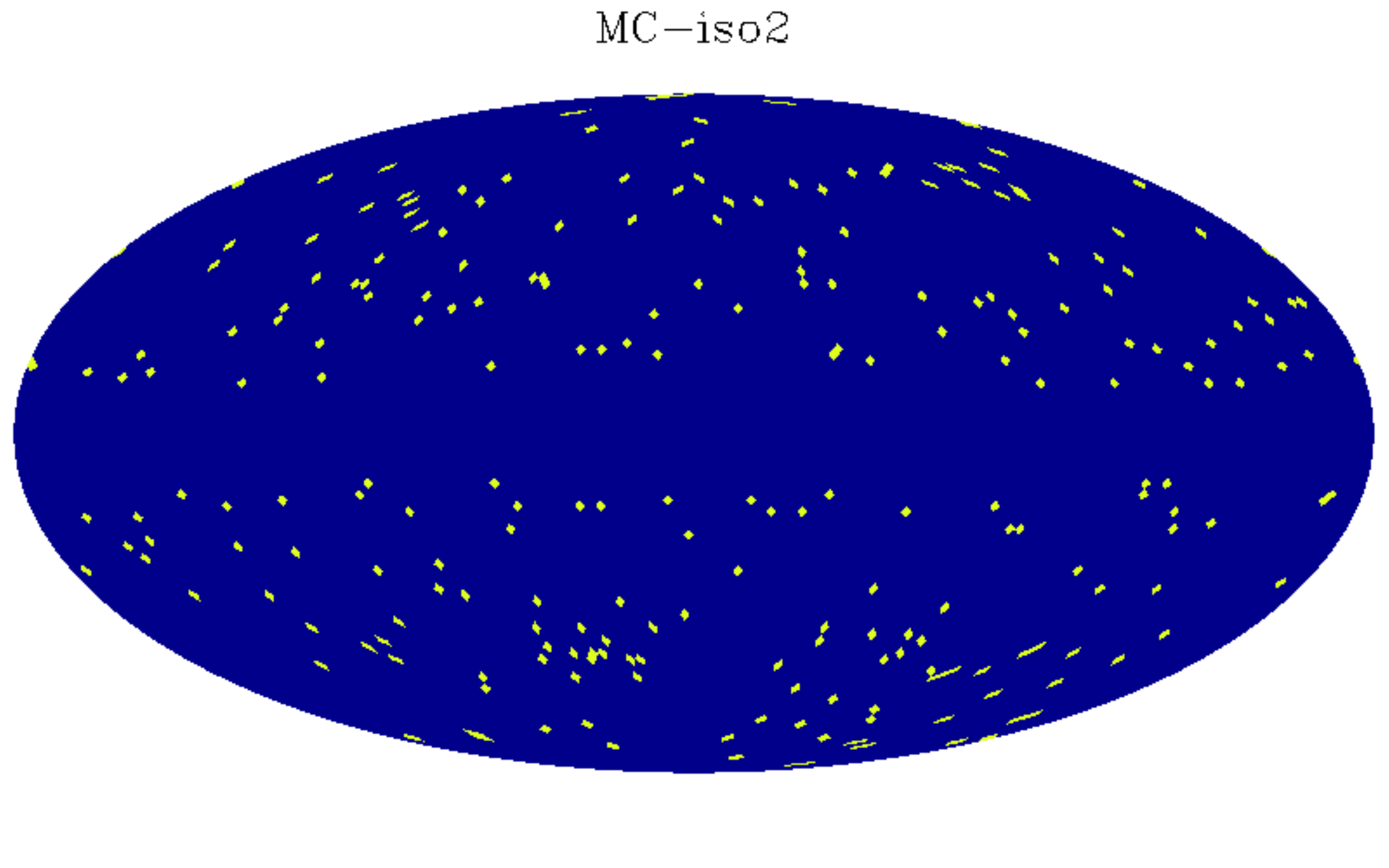}
\includegraphics[width=0.32\textwidth, height=0.215\textheight]{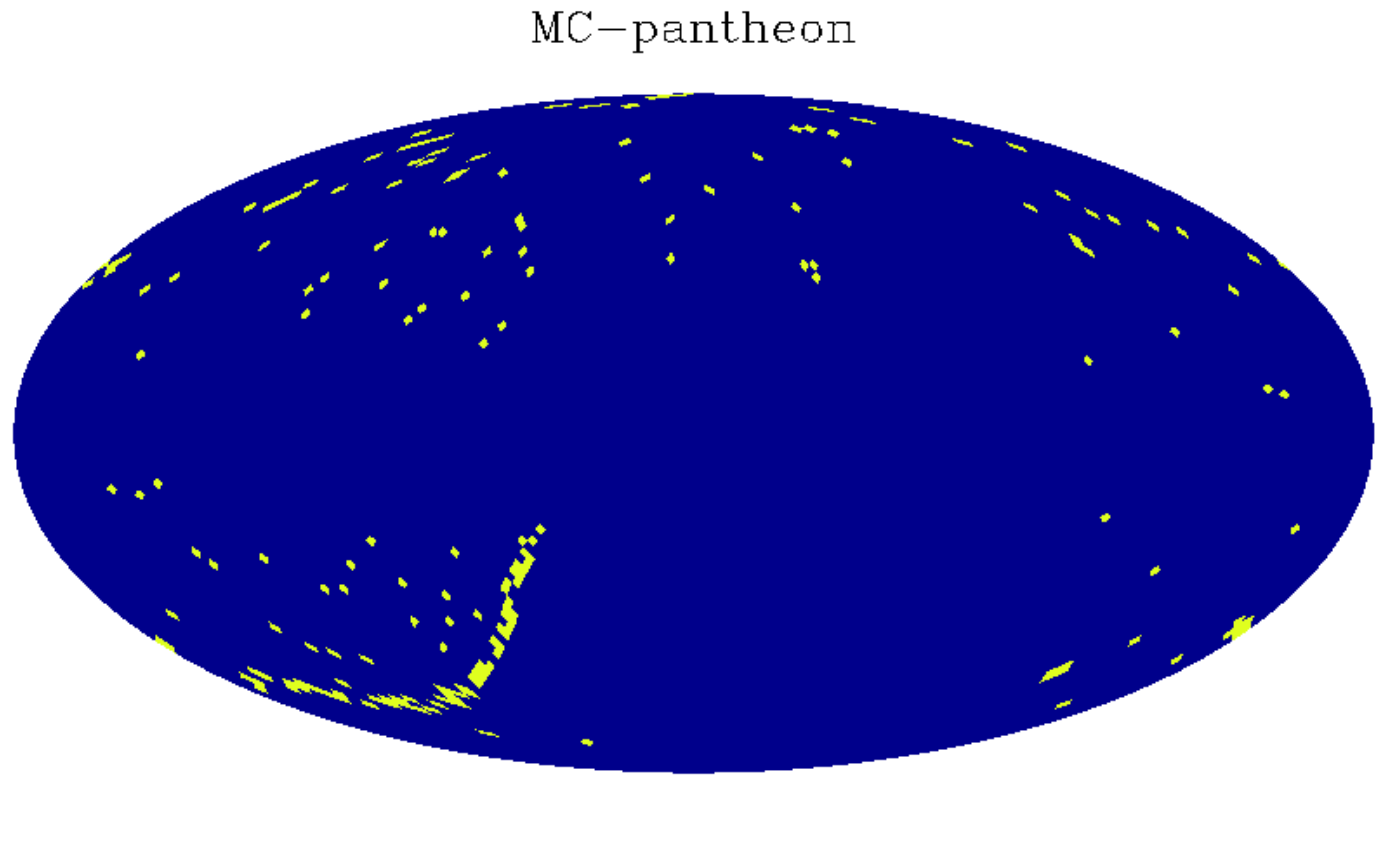}
\caption{Left panel: The sky distribution of SN at $0.023 < z < 0.150$ following the {\it MC-iso1} prescription. Central panel: Same as the left panel, but for the {\it MC-iso2} instead. Right panel: A realization assuming the original SN celestial distribution, corresponding to the {\it MC-pantheon} case.}
\label{fig:SN_maps}
\end{figure*}

In our analysis, we use the Pantheon SN Ia compilation, which consists of the currently  largest and most complete SN data-set, i.e., 1049 objects lying in the interval $0.01 < z < 2.30$ compiled from the PanSTARRS1 Medium Deep Survey, SDSS, SNLS, in addition to many low-$z$ and HST data points. As we will focus on a model-independent analysis, as explained further on with more details, we will only select objects in the range $0.023 < z  < 0.150$, hence reducing our sample to 237 data points. We exclude the SNe at $z<0.023$ to reduce the cosmic variance impact in very low redshift ranges (see~\cite{camarena18}). We do so by assuming that the FLRW metric holds true, thence one can expand the scale factor around the present time, and then measure distances regardless of the Universe dynamics. This is the well-known cosmographic approach largely discussed e.g. in~\cite{weinberg72, visser04}. The luminosity distance reads
\begin{eqnarray}\label{eq:DL_y}
D_{\rm L}(y)  = 3000h^{-1}_0 \left[y + \frac{(3 - q_0)}{2}{y^2} + O(y^3)\right] \;,
\end{eqnarray}
where $y \equiv z/(1+z)$~\citep{cattoen07} is the redshift observed in the comoving rest frame with respect to the expansion of the Universe, $h_0$ and $q_0$ are the dimensionless Hubble constant, defined as $h_0 \equiv H_0/100$, and decelerating parameter at present time, respectively, for $D_{\rm L}(y)$ given in Mpc. From $D_{\rm L}(y)$, we can obtain the distance modulus of the SN such as
\begin{eqnarray}\label{eq:mu}
\mu(y) = 5\log_{10}\left({D_{\rm L}(y)/\rm{Mpc}}\right) + 25 \;.
\end{eqnarray}
As shown in Eq.~\ref{eq:DL_y}, the luminosity distance depends only on $h_0$ and $q_0$ up to the second order in redshift. Therefore, we restrict our analysis up to that order. As shown in~\cite{bengaly15}, this truncation does not bias $\mu$ in the interval $z \lesssim 0.20$.  

We probe the $h_0$ cosmic variance by mapping the directional dependence of $h_0$ in the interval $0.023 < z < 0.150$, accordingly to the analysis of~\cite{bengaly16}. We define hemispheres whose symmetry axes are given by {\sc HEALPix}~\citep{gorski05} pixel centers at $N_{\rm side}=16$ grid resolution, and then we obtain $h_0$ best-fits for all SN enclosed in $3072$ distinct hemispheres. To do so, we minimize the following quantity
\begin{equation}\label{eq:chi2}
\chi^2 = \sum_{\rm i} \left[ ( \mu^{\rm th}(\mathbf{p},y_{\rm i})-\mu^{\rm obs}_{\rm i} ) \over \sigma_{\rm i} \right]^2 \,;
\end{equation}
where $\mathbf{p}$ represents the set of parameters $\{h_0,q_0\}$. The latter is fixed at $q_0 = -0.574$, thus fully consistent with  the Pantheon best fit for a flat $\Lambda$CDM model with $\Omega_{\rm m}=0.274$. A similar procedure was adopted in~\cite{kalus13}\footnote{We note that changing the value of the deceleration parameter within a reasonable interval (that includes the value $q_0 = -0.574$ adopted in the analysis) does not appreciably change our results.}. In addition, ${\rm{i}}$ represents the ${\rm{i}}$-th data point belonging to each hemisphere, $\mu^{\rm obs}_{\rm i}$ and $\sigma^2_{\rm i}$ correspond to its distance modulus and respective uncertainty, and $\mu^{\rm th}(\mathbf{p},y_{\rm i})$ represents the theoretically expected distance modulus calculated according to Eq.~(\ref{eq:mu}). We note that only the statistical errors for $\mu^{\rm obs}_{\rm i}$ are used in our analyses, as the full covariance matrix would significantly degrade the constraints at such low-$z$ range - specially because we are selecting hemispherical sub-samples of the SN data-set. Because our results are weakly sensitive to the cosmological model, we expect it to not be strongly affected by the SN light-curve nuisance parameters, whose values were also fixed at the flat $\Lambda$CDM model best-fits. 

The cosmic variance of the Hubble Constant is estimated from three sets of 1000 Monte Carlo (MC) realizations according to the following prescriptions: 

\begin{itemize}

\item {\it MC-iso1}: The SN original positions in the sky are changed according to an isotropic distribution, but with the galactic plane ($|\mathrm{b}|<10^{\circ}$), and the highest declinations ($\mathrm{DEC}>30^{\circ}$) excised;

\item {\it MC-iso2}: The SN original positions in the sky are changed according to an isotropic distribution, but with the galactic plane ($|\mathrm{b}|<10^{\circ}$) excised only;

\item {\it MC-pantheon}: The SN original positions are maintained as the original Pantheon sample. 

\end{itemize}

In all these realizations (see Fig. 1), the original distance moduli are changed according to a value drawn from a normal distribution, i.e., 
\begin{equation}\label{eq:mu_MC}
\mu^{\rm MC}_{\rm th}(y_{\rm i}) = \mathcal{N}(\mu_{\rm fid}(\mathbf{p_{\rm fid}},y_{\rm i}),\sigma_i) \,, 
\end{equation}
\noindent which corresponds to a distribution centered at $\mu_{\rm fid}$, that is, the distance modulus fixed at the fiducial Cosmology in $\mathbf{p_{\rm fid}} = \{h_{0;\mathrm{fid}},q_{0;\mathrm{fid}}\} = \{0.6736,-0.574$\}, and whose standard deviation is given by the original distance modulus uncertainty, $\sigma$, for each data point $i$. This way, we circumvent the problem of $h_0$ being implicitly assumed in the SN light-curve calibration procedure.

\begin{figure*}[!t]
\centering
\includegraphics[scale=0.45]{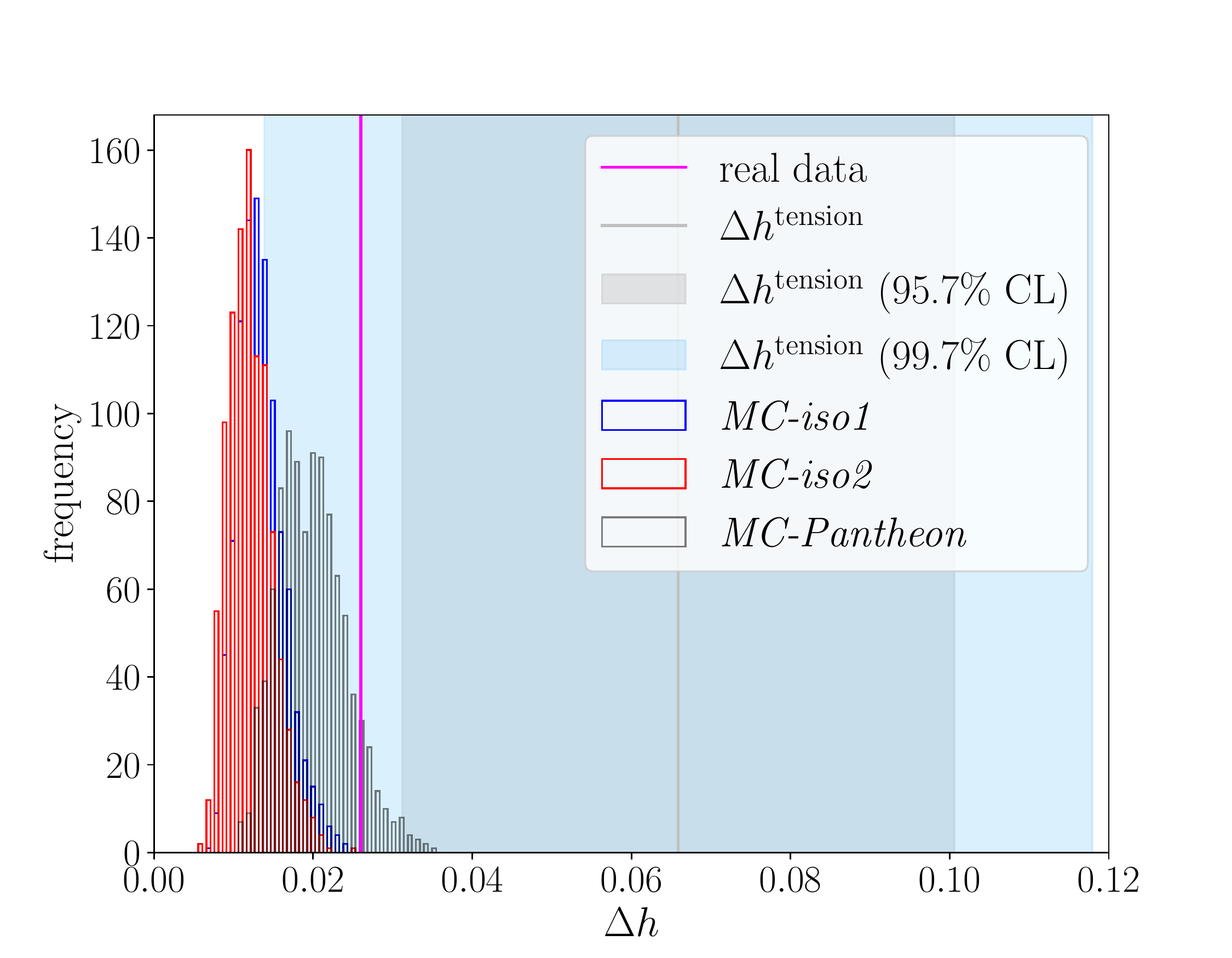}
\caption{The $\Delta h$ of all MC realizations compared to the value obtained from the real data (pink vertical line), as well as the $\Delta h^{\rm tension}$ at $1$ and $2\sigma$ C.L. displayed in light gray and blue shades, respectively.} 
\label{fig:deltah_pantheon}
\end{figure*}

\begin{figure*}[!t]
\centering
\includegraphics[width=0.49\textwidth, height=0.32\textheight]{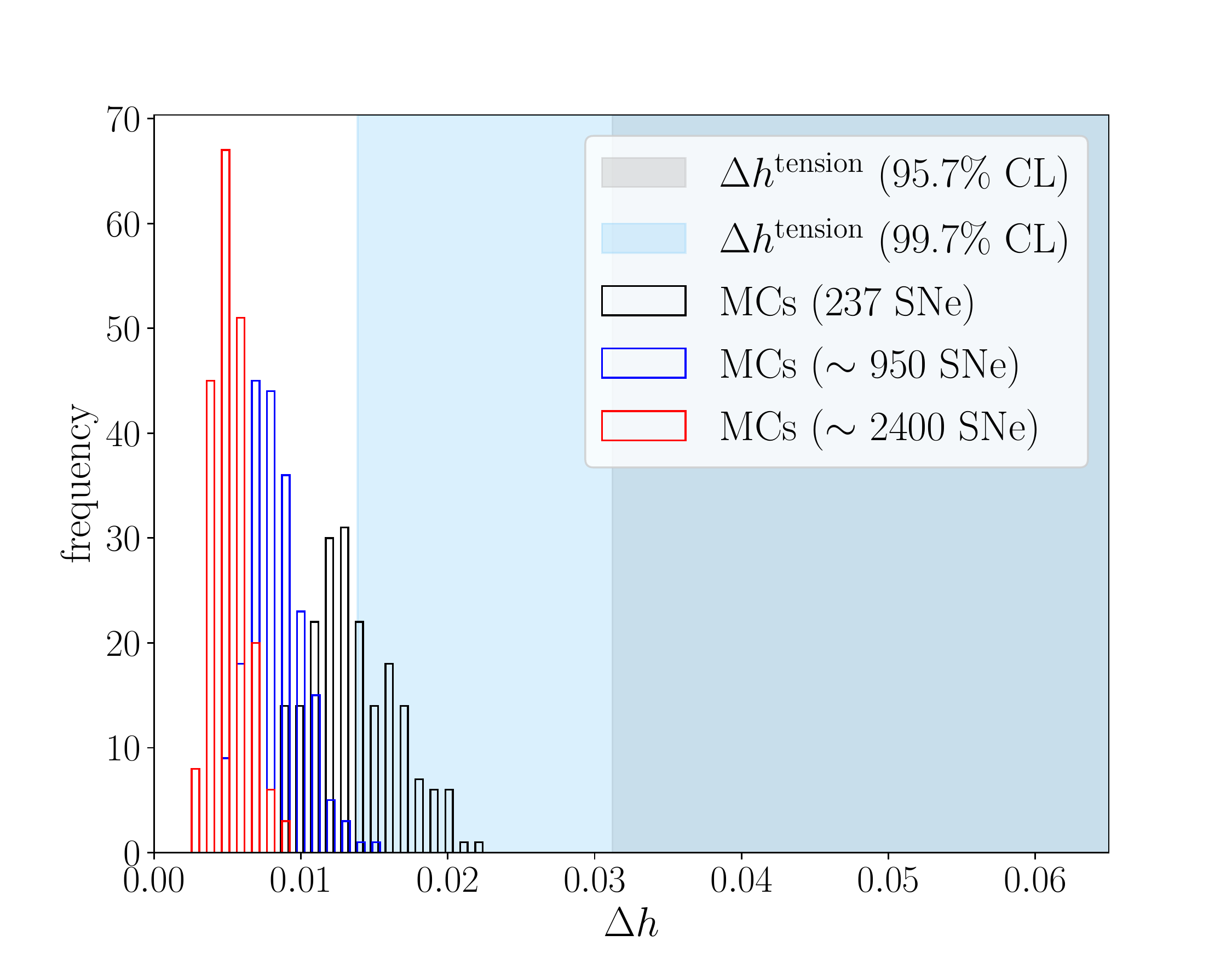}
\includegraphics[width=0.49\textwidth, height=0.32\textheight]{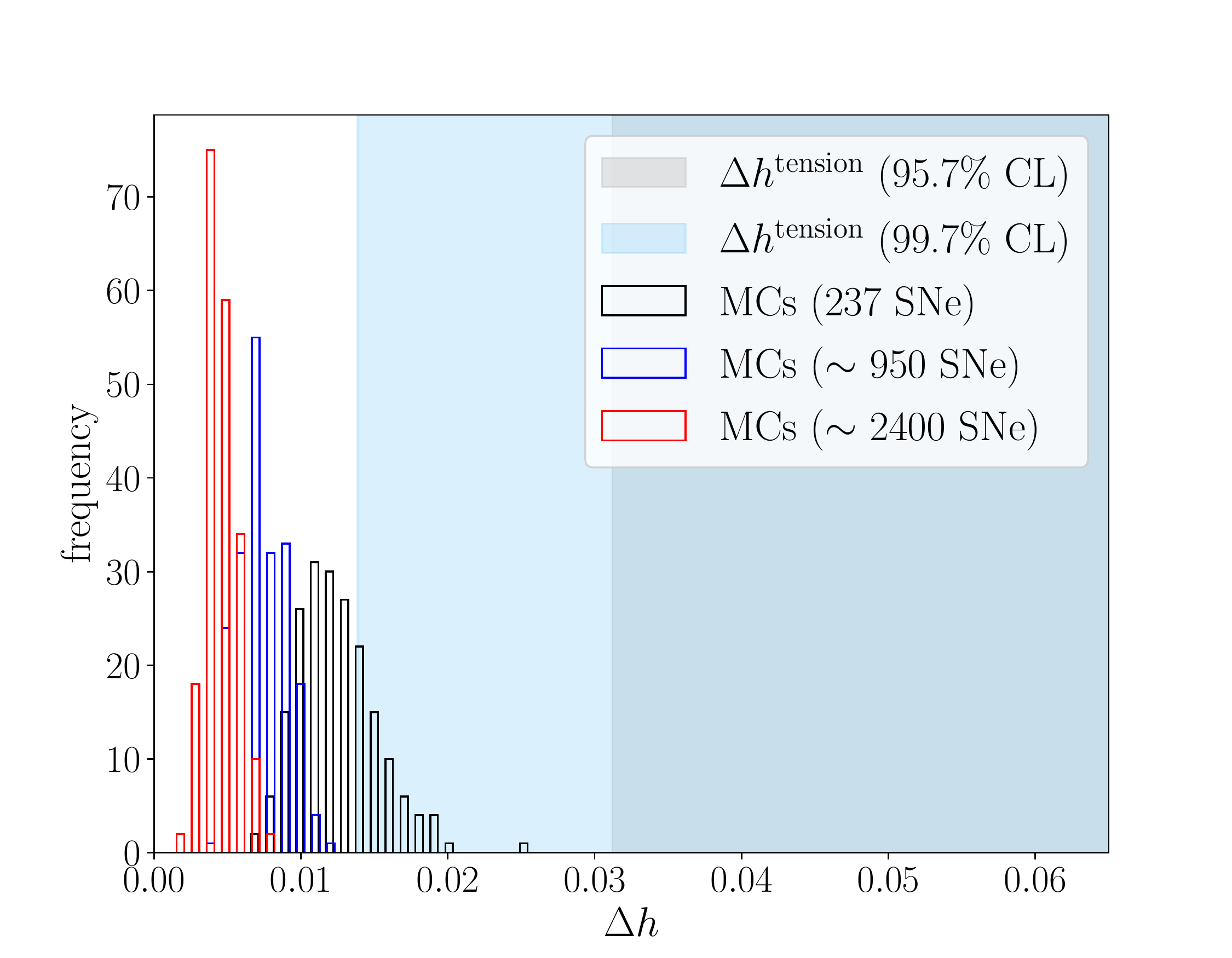}
\caption{Left panel: The $\Delta h$ of 200 {\it MC-iso1} realizations assuming the original number of data points, as well as for those assuming samples 4 and 10 times larger. Right panel: Same as before, but for {\it MC-iso2} instead. The light gray and blue shades are the same as Fig.~\ref{fig:deltah_pantheon}, but $\Delta h$ was clipped at $0.05$ to ease visualization.} 
\label{fig:deltah_mc}
\end{figure*}

We quantify the cosmic variance of the MCs by means of  
\begin{equation}\label{eq:delta}
\Delta h \equiv h^{\rm max}_0 - h^{\rm min}_0 \;,
\end{equation}
\noindent in a similar fashion to~\citep{bengaly16}, where $h^{\rm max}_0$ and $h^{\rm min}_0$ correspond to the maximum and minimum $h_0$ values fitted across the whole celestial sphere in each realization following Eq.~(\ref{eq:chi2}), where $\mu^{\rm MC}_{\rm th}(y_{\rm i}) \rightarrow \mu^{\rm obs}_{\rm i}$. For the sake of comparison, the tension can be quantified by
\begin{equation}\label{eq:delta_tension} 
T_{h_0} \equiv \frac{h^{\rm R19}_0 - h^{\rm P18}_0}{\sqrt{\sigma_{R19}^2 + \sigma_{P18}^2}} \simeq 4.4\sigma \;,
\end{equation}
where 
$\Delta h^{\rm tension} \equiv h^{\rm R19}_0 - h^{\rm P18}_0 = 0.0667$ and $\sigma_{\mathrm{R19}}^2 + \sigma_{\mathrm{P18}}^2 = 2.3 \times 10^{-4}$. 
We redefine the tension between the R19 and P18 measurements due to the cosmic variance similarly to~\cite{camarena18}, that is,
\begin{equation}\label{eq:T_h0}
T_{h_0} = \frac{h^{\rm R19}_0 - h^{\rm P18}_0}{\sqrt{\sigma_{\mathrm{R19}}^2 + \sigma_{\mathrm{P18}}^2 + \Delta h_{\rm med}^2}} \,,
\end{equation}
with $\Delta h_{\rm med}^2$ representing the median $\Delta h$ value for each MC case. Since we have a different $\Delta h$ value for each MC realization, we define the maximum and minimum value for $T_{h_0}$ by means of the upper ($\Delta h_{+}$) and lower ($\Delta h_{-}$) bounds of $\Delta h$ at 99.7\% CL such as
\begin{subequations}
\begin{equation}\label{eq:T+_h0}
T^{+}_{h_0} \equiv \frac{h^{\rm R19}_0 - h^{\rm P18}_0}{\sqrt{\sigma_{\mathrm{R19}}^2 + \sigma_{\mathrm{P18}}^2 + \Delta h_{+}^2}} \;,
\end{equation}
\begin{equation}\label{eq:T-_h0}
T^{-}_{h_0} \equiv \frac{h^{\rm R19}_0 - h^{\rm P18}_0}{\sqrt{\sigma_{\mathrm{R19}}^2 + \sigma_{\mathrm{P18}}^2 + \Delta h_{-}^2}} \,.
\end{equation}
\end{subequations}

Finally, we repeat these analyses for simulated (isotropic) data-sets 4 and 10 times larger than the current one assuming the same redshift distribution, and the same relative distance uncertainties as the real data. Hence, we estimate how well future data can constrain the $h_0$ cosmic variance and discuss its cosmological implications.


\section{Results}\label{sec:results}

\begin{table}[t]
\begin{center}
\begin{tabular}{ccccccc}
\hline
\hline
Case & $\Delta h_{\rm med}$ & $\Delta h$ (99.7\% CL) & $T_{h_0}$ & $T^{-}_{h_0}$ & $T^{+}_{h_0}$ \\
\hline
{\it MC-iso1}     & $0.013$ & $[0.008,0.022]$ & $3.336$ & $3.885$ & $2.495$ \\
{\it MC-iso2}     & $0.012$ & $[0.007,0.021]$ & $3.445$ & $3.988$ & $2.573$ \\
{\it MC-pantheon} & $0.020$ & $[0.012,0.031]$ & $2.656$ & $3.445$ & $1.932$ \\
\hline
\hline
\end{tabular}
\end{center}
\caption{Respectively: The MC prescription, the median $\Delta h$, the 99.7\% CL limits for $\Delta h$ obtained from each case, and the $h_0$ tension as defined in Eqs.~(\ref{eq:T_h0})-(\ref{eq:T-_h0}).}
\label{tab:tab_deltah_pantheon} 
\end{table}

\begin{table}
\begin{center}
\begin{tabular}{ccccccc}
\hline
\hline
Case ($\simeq$ 950 SN) & $\Delta h^{\rm med}$ & $\Delta h$ (99.7\% CL) & $T_{h_0}$ & $T^{-}_{h_0}$ & $T^{+}_{h_0}$ \\
\hline
{\it MC-iso1} & $0.008$ & $[0.005,0.014]$ & $3.885$ & $4.170$ & $3.229$ \\
{\it MC-iso2} & $0.007$ & $[0.005,0.011]$ & $3.988$ & $4.170$ & $3.556$ \\
\hline
\hline
Case ($\simeq$ 2400 SN) & $\Delta h^{\rm med}$ & $\Delta h$ (99.7\% CL) & $T_{h_0}$ & $T^{-}_{h_0}$ & $T^{+}_{h_0}$ \\
\hline
{\it MC-iso1} & $0.005$ & $[0.004,0.009]$ & $4.170$ & $4.246$ & $3.777$ \\
{\it MC-iso2} & $0.005$ & $[0.002,0.007]$ & $4.170$ & $4.352$ & $3.988$ \\
\hline
\hline
\end{tabular}
\end{center}
\caption{Respectively: Same as Table~\ref{tab:tab_deltah_pantheon}, but for 200 MCs assuming 4 and 10 times more data points than the current sample instead.}
\label{tab:tab_deltah_mocks} 
\end{table}

In Fig.~\ref{fig:deltah_pantheon}, we show the results obtained from 1000 MCs with 237 data points for each sky coverage configuration. The blue histogram gives the $\Delta h$ distribution for the {\it MC-iso1} case, whereas the red and black ones  provide, respectively, the {\it MC-iso2} and {\it MC-pantheon} results. The light blue and gray shades correspond to the $\Delta h^{\rm tension}$ quantity for $2$ and $3\sigma$ CL, and the pink vertical line is the $\Delta h$ obtained from the actual Pantheon data as reported in~\cite{andrade18}. We note that the $\Delta h$ obtained from all the MCs overlap with this $3\sigma$ shade, but the {\it MC-pantheon} gives the largest values among all of them, illustrating that the non-uniform sky coverage increases the $h_0$ cosmic variance 

In Table~\ref{tab:tab_deltah_pantheon}, we provide the median, upper and lower values (at 99.7\% CL) for $\Delta h$, in addition to $T_{h_0}$ and their respective upper and lower limits. We readily note that the {\it MC-pantheon} realizations perform the best, as $T_{h_0}$ decreases from $\simeq 4.4\sigma$ to $2.7\sigma$ for the median $\Delta h$, and $1.9\sigma \, (3.4\sigma)$ for $T^{+}_{h_0}$ ($T^{-}_{h_0}$). Both isotropic sets of MCs can alleviate the tension to $3.4\sigma$ for the median case, and $2.5\sigma$ for the most optimistic ($T^{+}_{h_0}$) one. We also note that the median $\Delta h$ from both isotropic MCs are consistent with the estimates from~\cite{marra13, camarena18}, i.e., $\Delta h \simeq 0.009$, but in our case we can naturally incorporate other observational features of incomplete celestial coverage and SN distance measurement uncertainties, which contributes to reduce the tension even further. We also tested whether the individual uncertainties of the best-fitted $h_0$ values affect our analyses. We found that their average $1\sigma$ uncertainties for isotropic MCs lead to an additional (average) spread of $0.004$ in $\Delta h$ for the isotropic MCs, and $0.005$ for the MC-pantheon case. Since these values are smaller than the $\Delta h$ values presented in Table~\ref{tab:tab_deltah_pantheon}, we stress that they do not impact our analysis.

Moreover, we estimate the $h_0$ variance of future SN compilations. We do this by producing 200 {\it MC-iso1} and {\it MC-iso2} realizations according to the same redshift distribution of SN as the Pantheon sample for two cases: one assuming data-sets 4 times larger than the Pantheon sample (nearly 950 objects), and another assuming 10 times its size (nearly 2400 objects). If $\Delta h$ reduces in these cases, this would imply that cosmic variance cannot alleviate the $\Delta h^{\rm tension}$ - and hence increase the evidence for physics beyond $\Lambda$CDM as its explanation.  

We present the results of these analyses in Figure~\ref{fig:deltah_mc}. The $\Delta h$ values and $T_{h_0}$ are provided in table~\ref{tab:tab_deltah_mocks}. We can clearly note that these $\Delta h$ are much smaller compared to the previous sets of MCs, so that $T_{h_0} > 3.2\sigma$ for the realizations with 950 data points, and $T_{h_0} > 3.8\sigma$ for those with 2400 objects. Given the advent of forthcoming, nearly all-sky distance data-sets as expected from LSST~\citep{lsst09} and WALLABY~\citep{duffy12}, along with future $H_0$ measurements from standard sirens~\citep{mortlock18, feeney19}, and new assessments of the cosmic distance ladder~\cite{verde19}, we should be able to determine whether the $H_0$ tension arises due to its cosmic variance from uneven sky coverage and limited observational data, or whether it actually indicates physics beyond the concordance model.

\section{Conclusions}\label{sec:conc}

The tension between the R19 and P18 estimates of $H_0$ are one of the greatest puzzles of the concordance model of Cosmology at present. Because the Hubble constant is fundamental to our understanding of the cosmic evolution, it is necessary to seek explanations for this tension. The main ones correspond to an additional error budget to the $H_0$ measurement due to local inhomogeneities and sample incompleteness, and to extensions of $\Lambda$CDM model such as time-evolving dark energy, interacting dark energy or new contributions to the effective number of relativistic species $N_{\rm eff}$, among others. None of them could fully account for the $H_0$ tension, but only alleviate it to $\simeq 3\sigma$. 

We focused on the former as a possible explanation for such a tension. This time around, we estimate the $H_0$ cosmic variance from a hemispherical comparison estimator, so we can quantify how does the incomplete data sampling affect the $H_0$ tension. We produced realizations of the Pantheon SN in the interval $0.023 < z < 0.150$ uniformly redistributed across the sky ({\it MC-iso1} and {\it MC-iso2}), besides those assuming the actual distribution of objects ({\it MC-pantheon}). We also replaced the actual SN distance moduli to values drawn from a normal distribution assuming the $\Lambda$CDM model, but the original distance errors. Hence, we quantified the $H_0$ variance due to the sample incompleteness within the concordance model framework.

We found that the $H_0$ tension could be alleviated from nearly $4.4\sigma$ to $2.5\sigma$, at best, for the isotropic simulations, but the MCs assuming the actual SN distribution gave a median value of $\simeq 2.7\sigma$ for a lower (upper) bound of $1.9\sigma \, (3.4\sigma)$. These results significantly improves previous analyses that pointed out $T_{h_0} \simeq 3\sigma$ due to the $H_0$ cosmic variance, as we could incorporate further SN sample limitations, such as uneven sky coverage and distance measurement uncertainties. Furthermore, we verified that the $\Delta h$ will be significantly smaller when larger samples becomes available, and so we can rule out cosmic variance as an explanation for this tension.

Our conclusion is that the cosmic variance can alleviate the $H_0$ tension - specially if the non-uniform celestial coverage is taken into account. It is worth mentioning that there might also be unidentified systematics in the observations that lead to such tension, as discussed in~\cite{bernal18}. As mentioned earlier, the forthcoming, nearly all-sky distance data make the prospects very good for solving this issue in the near future.

\emph{Acknowledgments} -- The authors thank Valerio Marra and the anonymous referee for useful discussions. UA acknowledges financial support from CAPES. CAPB acknowledges financial support from the South African SKA Project. JSA acknowledges support from CNPq (Grants no. 310790/2014-0 and 400471/2014-0) and FAPERJ (Grant no. 204282). Some of our analyses used the {\sc HEALPix} software package. 

%

\end{document}